\documentclass[aps,prb,twocolumn]{revtex4}
\usepackage{array,amsmath,amssymb,graphicx}
\usepackage{bm}

\begin{document}

\title{Biexcitons in two-dimensional systems with
spatially separated electrons and holes}

\author{A. D. Meyertholen and M. M. Fogler}


\affiliation{Department of Physics, University of
California San Diego, La Jolla, 9500 Gilman Drive, California 92093}

\date{\today}

\begin{abstract}

The binding energy and wavefunctions of two-dimensional indirect biexcitons are
studied analytically and numerically. It is proven that stable biexcitons exist
only when the distance between electron and hole layers is smaller than a
certain critical threshold. Numerical results for the biexciton binding energies
are obtained using the stochastic variational method and compared with the
analytical asymptotics. The threshold interlayer separation and its uncertainty
are estimated. The results are compared with those obtained by other techniques,
in particular, the diffusion Monte-Carlo method and the Born-Oppenheimer
approximation.

\end{abstract}

\maketitle

\section{The problem and main results}
\label{sec:Problem}

The physics of cold excitons --- bound states of electrons and holes
in semiconductors --- has attracted much attention recently. Cooling
the excitons has become possible by confining electrons and holes in
separate two-dimensional (2D) quantum wells, which greatly increases
their lifetime. A number of intriguing phenomena has been
demonstrated for such ``indirect'' excitons, including long-range
transport,~\cite{Hagn_95, Larionov_00, Voros_05, Butov_02,
Ivanov_06, High_08} macroscopic spatial ordering,~\cite{Butov_02}
and spontaneous coherence.~\cite{Yang_06} Theoretical work on these
phenomena is ongoing, see Ref.~\onlinecite{Butov_07} for review.
Further progress in this field requires an improved understanding of
exciton interactions.

Despite being charge neutral, indirect excitons possess a dipole moment $e d$,
where $d$ is the separation of the electron and hole quantum wells. As a result,
interaction of two excitons at large distances $r$ is
dominated by their dipolar repulsion,
\begin{equation} \label{eq:dipolar}
 V(r) = \frac{e^2}{\kappa} \frac{d^2}{r^3}\,,
\end{equation}
where $\kappa$ is the dielectric constant of the semiconductor. At short
distances exchange and correlation effects are also important. The interaction
may even become attractive over a range of $r$. In this case two excitons can
form a bound state --- a biexciton. The corresponding binding energy is defined
by
\begin{equation}
 E_{B} = 2 E_\text{X} - E_\text{XX}\,,
 \label{eq:E_B}
\end{equation}
where $E_\text{X}$ and $E_\text{XX}$ are the ground-state energies of
the exciton and biexciton, respectively.

\begin{figure}[b]
\centerline{
\includegraphics[width=2.9in]{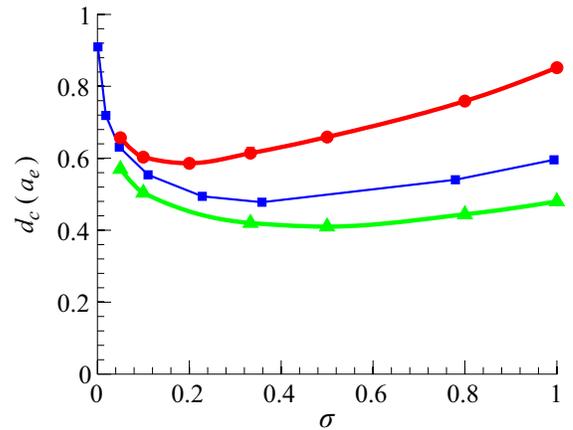}
}
\caption{Critical interlayer separation \textit{vs\/}. the electron-hole mass
ratio. The circles are our results. The squares are from
Ref.~\onlinecite{Zimmermann_08}. The triangles correspond to $d$ above which
$E_B(d)$ drops below $10^{-3}\,\text{Ry}_e$, making biexcitons irrelevant in
experimental practice.
\label{fig:d_c}}
\end{figure}

While observations of biexcitons in single quantum well structures
($d = 0$) have been described multiple times,~\cite{Kleinman_82,
Phillips_92, Bar-Ad_92, Birkedal_96, Adachi_04, Kim_98, Langbein_02,
Maute_03} no such reports exist for the $d > 0$ case. A recent
theoretical work~\cite{Tan_05} has attributed the lack of
experimental signatures of indirect biexcitons to extreme smallness
of their binding energies. In this paper we verify and improve all
previously known estimates of $E_B$. In particular, we show that
$E_{B}(d)$ is positive, i.e., the biexciton is stable, only for $d$
smaller than some critical value $d_c$, see Fig.~\ref{fig:d_c}.
Typical experimental parameters~\cite{Butov_07, Butov_98} fall on
the $d > d_c$ part of the diagram.

%
%
%

In our calculations we adopt the simplifying assumption that the
effective masses $m_e$ and $m_h \geq m_e$ of electrons and holes are
constant and isotropic. We also treat the quantum wells as 2D layers of
zero thickness. We find it convenient to measure distances in units of
the effective electron Bohr radius and energies in units of the
effective Rydberg,
\begin{equation} \label{eq:a_e}
           a_{e} = \frac{\hbar^2 \kappa}{m_e e^2}\,,
\quad  \text{Ry}_e = \frac12 \frac{e^2}{\kappa a_e}\,,
\end{equation}
respectively. With these conventions, the four-particle system of two electrons
and two holes is described by the Hamiltonian $H_\text{XX} = T + U$, where
\begin{gather}
\label{eq:T}
T = T_1 + T_2\,,\quad
T_j = -\nabla_j^2 - \sigma
       \left(\frac{d}{d \mathbf{R}_j}\right)^2\,,\\
\label{eq:U}
U = \frac{2}{|\mathbf{r}_1 - \mathbf{r}_2|}
  + \frac{2}{|\mathbf{R}_{1} - \mathbf{R}_{2}|}
  - \sum_{i j} v(\mathbf{r}_i - \mathbf{R}_{j}, d)\,,\\
\label{eq:v}
v(\mathbf{r}, d) =
       \frac{2}{\sqrt{|\mathbf{r}|^2 + d^2}}\,.
\end{gather}
Here $\mathbf{r}_i$ and $\mathbf{R}_i$ are 2D coordinates of the electrons and
the holes, respectively, $\nabla_j = d / d \mathbf{r}_j$, and
\begin{equation} \label{eq:sigma}
                  \sigma = {m_e} / {m_h}
\end{equation}
is the mass ratio. Similarly, the single-exciton Hamiltonian is
\begin{equation} \label{eq:H_X}
        H_X = T_1 + v(\mathbf{r}_1 - \mathbf{R}_{1}, d)\,.
\end{equation}
The problem is characterized by two dimensionless parameters: $d$ and
$\sigma$. The case of $d = 0$ (direct excitons) has been studied
extensively.~\cite{Bressanini_98, Usukura_99, Riva_02} In contrast, high
accuracy calculations of $E_{B}$ for $d > 0$ have been carried out only
in the aforementioned Ref.~\onlinecite{Tan_05}. The authors of that work
have employed the diffusion quantum Monte-Carlo method (DMC). Away from
$d = 0$, they were able to fit their results for $\sigma = 1$ and
$\sigma = 1 / 2$ to the exponential:
\begin{equation} \label{eq:Tan}
 E_{B}(d) \approx \alpha e^{-\beta d}.
\end{equation}
This result is surprising. Equation~\eqref{eq:Tan} seems to
imply that the biexcitons are stable at any $d$, i.e., $d_c = \infty$.
On the other hand, physical intuition and previous approximate
calculations~\cite{Lozovik_97, Laikhtman_02} suggest that $d_c$ should
be finite. A more recent work~\cite{Zimmermann_08} has reached the same
conclusion. In this paper we present rigorous analytical
arguments and essentially exact numerical results proving that $d_c
\leq 1$ at all $\sigma$, see Fig.~\ref{fig:d_c}. (Due to
electron-hole symmetry, it is sufficient to consider $0 \leq \sigma \leq
1$.)

Since $d_c$ is finite, the interpolation formula~\eqref{eq:Tan} must
overestimate the binding energy at large $d$. We show that near the
biexciton dissociation threshold,
\begin{equation}
 \label{eq:asym_range}
                     d_c - d \ll D\,,
\end{equation}
where $D \sim 1$ for $\sigma \sim 1$ and $D \sim \exp(-\sigma^{-1/2})$ for
$\sigma \ll 1$, function $E_{B}(d)$ behaves as
\begin{equation}
 \label{eq:E_B_asym}
 E_{B} \simeq E_{0} e^{-{D} / {(d_c - d)}}.
\end{equation}
This equation resembles the well-known expression for the energy
$\varepsilon$ of a bound state in a weak 2D potential $V(r)$. Such a
state exists if
\begin{equation}
 \label{eq:W}
W \equiv \frac{M}{2 \pi \hbar^2} \int d^2 r V(r) < 0\,,
\end{equation}
where $M$ is the mass of the particle. Near the threshold $W \to 0$
one finds~\cite{Landau_III}
\begin{equation}
 \label{eq:Landau}
 |\varepsilon| \propto e^{-1 / |W|}, \quad |W| \ll 1\,.
\end{equation}
The exciton-exciton interaction potential $V(r)$ in general does not
satisfy the condition of the perturbation theory $V(r) r^2 \ll
\hbar^2 / M$, with $M = m_e + m_h$. Therefore, Eq.~\eqref{eq:Landau}
does not literally apply here. Nevertheless, the physical origins of
the exponential dependence in Eqs.~\eqref{eq:E_B_asym} and
\eqref{eq:Landau} are the same, see Sec.~\ref{sec:Analytical}B.

\begin{figure}[t]
\centerline{
\includegraphics[width=2.9in]{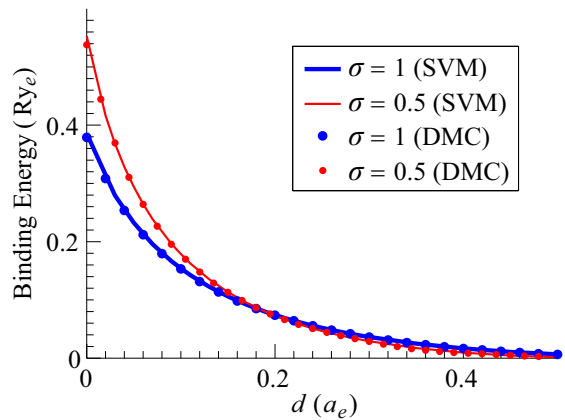}
}
\caption{Binding energy \text{vs\/}. the distance between the quantum wells for
the mass ratios $\sigma = 1$ and $0.5$. Our results are shown by
the solid lines. The dots are from the Ref.~\onlinecite{Tan_05}.
\label{fig:E_B}
}
\end{figure}

We verify and complement the above analytical results numerically using
the stochastic variational method (SVM).~\cite{Varga_95} The SVM has
proven to be a powerful technique for computing the energies of
few-particle systems.~\cite{SVM_book} For example, it has given the best
estimates of $E_B$ for direct biexcitons,~\cite{Bressanini_98,
Usukura_99} $d = 0$. Our calculations are largely in excellent agreement
with those of Ref.~\onlinecite{Tan_05}, see Fig.~\ref{fig:E_B} and
Table~\ref{table:E_B}. Thus, Eq.~\eqref{eq:Tan} is certainly useful as
an interpolation formula for not too large $d$. However, near the estimated
$d_c$, our results favor Eq.~\eqref{eq:E_B_asym} over
Eq.~\eqref{eq:Tan}. Since the SVM is variational, we can be sure that it
is more reliable when it gives a larger $E_{B}$ than other methods.
%

%
%
%

The remainder of the paper is organized as follows. In Sec.~\ref{sec:Analytical}
we derive a few analytical bounds on $E_B$ and the asymptotic
formula~\eqref{eq:E_B_asym}. Numerical calculations are presented in
Sec.~\ref{sec:Numerical}. Section~\ref{sec:Discussion} is devoted to discussion
and comparison with results in previous literature. Some details of the derivation
are given in the Appendix.

\begin{table}[b]
\caption{Biexciton binding energies in units of $\text{Ry}_e$ from
the previous (``DMC'', Ref.~\onlinecite{Tan_05}) and
present (``SVM'') work.
\label{table:E_B}
}
\begin{tabular}{c@{\hspace{0.2in}}%
                c@{\hspace{0.2in}}%
                c@{\hspace{0.4in}}%
                c@{\hspace{0.2in}}%
                c@{\hspace{0.2in}}%
                c}
\hline \hline
       &$\sigma = 1$&      &      &$\sigma = 0.5$&   \\
$d(a_e)$& DMC   & SVM      &$d(a_e)$& DMC  & SVM     \\
\hline
  0.00 & 0.3789 & 0.3858   & 0.00 & 0.5381 & 0.5526  \\
  0.02 & 0.3084 & 0.3089   & 0.01 & 0.4443 & 0.4450  \\
  0.04 & 0.2538 & 0.2546   & 0.03 & 0.3695 & 0.3689  \\
  0.06 & 0.2118 & 0.2133   & 0.04 & 0.3104 & 0.3109  \\
  0.08 & 0.1794 & 0.1807   & 0.06 & 0.2639 & 0.2649  \\
  0.10 & 0.1532 & 0.1542   & 0.07 & 0.2265 & 0.2275  \\
  0.12 & 0.1315 & 0.1324   & 0.09 & 0.1956 & 0.1966  \\
  0.14 & 0.1135 & 0.1141   & 0.11 & 0.1696 & 0.1707  \\
  0.16 & 0.0982 & 0.0986   & 0.12 & 0.1477 & 0.1487  \\
  0.18 & 0.0851 & 0.0855   & 0.14 & 0.1291 & 0.1299  \\
  0.20 & 0.0738 & 0.0742   & 0.15 & 0.1130 & 0.1136  \\
  0.22 & 0.0640 & 0.0644   & 0.17 & 0.0989 & 0.0995  \\
  0.24 & 0.0556 & 0.0559   & 0.18 & 0.0865 & 0.0872  \\
  0.26 & 0.0483 & 0.0485   & 0.20 & 0.0757 & 0.0764  \\
  0.28 & 0.0418 & 0.0420   & 0.21 & 0.0663 & 0.0670  \\
  0.30 & 0.0361 & 0.0363   & 0.23 & 0.0580 & 0.0586  \\
  0.32 & 0.0311 & 0.0313   & 0.24 & 0.0507 & 0.0512  \\
  0.34 & 0.0267 & 0.0270   & 0.26 & 0.0443 & 0.0447  \\
  0.36 & 0.0229 & 0.0231   & 0.27 & 0.0385 & 0.0389  \\
  0.38 & 0.0195 & 0.0197   & 0.28 & 0.0333 & 0.0337  \\
  0.40 & 0.0165 & 0.0167   & 0.30 & 0.0286 & 0.0291  \\
  0.42 & 0.0140 & 0.0141   & 0.32 & 0.0241 & 0.0250  \\
  0.44 & 0.0117 & 0.0118   & 0.33 & 0.0200 & 0.0214  \\
  0.46 & 0.0096 & 0.0097   & 0.34 & 0.0165 & 0.0182  \\
  0.48 & 0.0078 & 0.0079   & 0.36 & 0.0135 & 0.0154  \\
  0.50 & 0.0063 & 0.0065   & 0.38 & 0.0112 & 0.0129  \\
  0.52 & 0.0051 & 0.0052   & 0.39 & 0.0096 & 0.0107  \\
  0.54 & 0.0040 & 0.0040   & 0.41 & 0.0087 & 0.0087  \\
  0.56 & 0.0030 & 0.0031   & 0.42 & 0.0076 & 0.0071  \\
  0.58 & 0.0021 & 0.0023   & 0.44 & 0.0064 & 0.0056  \\
  0.60 & 0.0013 & 0.0017   & 0.45 & 0.0051 & 0.0044  \\
  0.62 & 0.0007 & 0.0012   & 0.47 & 0.0039 & 0.0033  \\
  0.64 & 0.0002 & 0.0007   & 0.48 & 0.0027 & 0.0024  \\
\hline \hline
\end{tabular}
\end{table}

\section{Analytical results}
\label{sec:Analytical}

In this section we approach the biexciton problem by analytical methods. Since
the exact solution seems out of reach, the best one can do is to consider
certain limits where suitable control parameters exist. Below we examine three
of them. First, we study large-$d$ excitons. We prove that they cannot bind into
a stable biexciton. Second, we consider the immediate vicinity $d_c - d \ll 1$
of the dissociation threshold $d_c$. We derive the asymptotical formula for the
binding energy, Eq.~\eqref{eq:E_B_asym}, which is valid for arbitrary $\sigma$.
Finally, we analyze the case $\sigma \ll 1$.

\subsection{Exciton interaction at large $\bm{d}$}

The absence of stable biexcitons at large $d$ is due to the
lack of binding in the classical limit, which is realized at such $d$.
Indeed, if we temporarily change the length units to $d$ and energy
units to $e^2 / \kappa d$, then the potential energy $U$ in
Eq.~\eqref{eq:U} becomes $d$-independent while the kinetic energy $T$
acquires the extra factor $a_e / d \ll 1$ compared to Eq.~\eqref{eq:T}.
Hence, the potential energy dominates. A rigorous proof that $d_c <
\infty$ can be constructed by dealing with the quantum and many-body
aspects of the problem separately. The many-body part is handled at the
classical level. Thereafter the quantum corrections are included. With
further analysis, both parts of the argument can be reduced to simpler
problems for which controlled approximations exist.

%
%
%
\begin{figure}[t]
\centerline{
\includegraphics[width=3.0in]{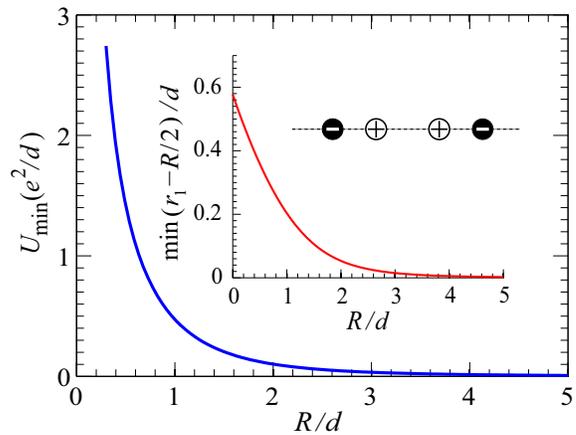}
}
\caption{Main panel: ground-state energy $U_{\min}$ \textit{vs\/}. the
separation $R$ of holes for a pair of classical excitons. In this state all four
charges are on the same straight line. Inset: in-plane distance between nearest
electrons and holes \textit{vs\/}. $R$.
\label{fig:classical}}
\end{figure}

Since the Earnshaw theorem does not apply in 2D, the absence of a stable
classical biexciton is not immediately obvious. However, we verified it
following these steps. The classical ground-state is the global minimum
of the potential energy. We can do the minimization over the electron
positions $\mathbf{r}_1$
and $\mathbf{r}_2$ first. Let $\mathbf{R}$ be the distance between the
holes,
\begin{equation}
\label{eq:R}
\mathbf{R} = \mathbf{R}_1 - \mathbf{R}_2\,,
\end{equation}
then the energy function to minimize is (in the original units convention)
\begin{equation}
\label{eq:U_R}
U_R = \frac{2}{|\mathbf{r}_1 - \mathbf{r}_2|} + \frac{2}{R}
- \sum_{\substack{j = 1, 2 \\ \mathbf{t} = \pm \mathbf{R} / 2}}
                             v(\mathbf{r}_j - \mathbf{t}, d)\,.
\end{equation}
It can be shown that for all $R$ the lowest energy is achieved when
the in-plane coordinates of the four charges fall on a straight
line, see Fig.~\ref{fig:classical}. Forming a cross is the only
other viable alternative, but it always has a higher energy. For the
linear geometry of the system, numerically exact results for
$U_{\min}(R, d) \equiv \min_{\mathbf{r}_1, \mathbf{r}_2} U_R$ are
obtained trivially. The plot of $V_{\text{cl}}(R) \equiv U_{\min}(R,
d) + (4 / d)$ is shown in Fig.~\ref{fig:classical}. This combination
can be thought of as the classical limit of the exciton interaction
potential $V(R)$. Function $V_{\text{cl}}$ monotonously decreases
with $R$ and achieve its global minimum at $R = \infty$. This means
that classical excitons do not form a bound state.

At large $R$, function $V_{\text{cl}}(R)$ follows the dipolar
interaction law~\eqref{eq:dipolar} with the quadrupolar,
\textit{etc.\/}, corrections:
\begin{equation}
\label{eq:V_cl}
  V_{\text{cl}}(R, d) = \frac{2 d^2}{R^3}
                      - \frac32 \frac{d^4}{R^5}
                      + \mathcal{O}\left(\frac{d^6}{R^7}\right)\,,
\quad R \gg d\,.
\end{equation}
Quantum corrections due to the zero-point motion about the classical
ground state are not able to compete with the dipolar repulsion when $d$
is large, see Appendix~\ref{sec:Proof}. Therefore, there is a critical
$d_c = d_c(\sigma)$ above which a stable biexciton does not exist.

\subsection{Binding energy near $\bm{d_c}$}

In this subsection we examine the biexciton state near the
dissociation threshold $d_c$ for arbitrary $\sigma$. It is easy to
understand that in this regime the biexciton orbital wavefunction
$\Psi$ should have a long tail extending to large distances away
from the center of mass of the the system. Inside of this tail the
configurations of electrons and holes resemble a pair of
well-separated individual excitons. Therefore, at $r \gg 1$, where
$r$ is the distance between the centers of mass of two such
excitons, $\Psi$ takes the asymptotic form
\begin{align}
\label{eq:Phi}
\Psi &= \left[1 + (-1)^s P_{1 2}\right] \Phi(\mathbf{r})
       \prod_{j = 1, 2} \phi_\sigma(\mathbf{r}_j - \mathbf{R}_{j})\,,
\\
\label{eq:r}
\mathbf{r} &= \frac{1}{1 + \sigma} \mathbf{R}
            + \frac{\sigma}{1 + \sigma} (\mathbf{r}_{1} - \mathbf{r}_{2})
\,.
\end{align}
Here $s$ is the total electron spin, $\phi_\sigma$ is the ground-state
wavefunction of a single exciton with mass ratio $\sigma$, and operator $P_{1
2}$ exchanges $\mathbf{r}_1$ and $\mathbf{r}_2$. Let us assume, for simplicity,
that holes are spin-$1 / 2$ particles. Then the wavefunction $\Phi$ of the
relative motion must have the parity $\Phi(-\mathbf{r}) = (-1)^{s + S}
\Phi(\mathbf{r})$, where $S$ is the total spin of the holes. Our goal in this
subsection is to determine the behavior of $\Phi$ at large $r$ and use it to
derive Eq.~\eqref{eq:E_B_asym}.

We proceed, as usual, by expanding $\Phi$ into partial waves of
angular momenta $m$ ($m$ and $s + S$ must be simultaneously odd or
even). The equation for the radial wavefunction $\chi_m(r)$ reads
\begin{equation} \label{eq:chi_m}
 -\frac{1}{r} \frac{d}{d r} r \frac{d \chi_m}{d r} +
   \left[
   \varkappa^2 + \mu V(r) + \frac{m^2}{r^2}
   \right] \chi_m = 0\,,
\end{equation}
where $\varkappa$ and $\mu$ are defined by
\begin{equation} \label{eq:kappa}
\varkappa = \sqrt{\mu E_B}\,,
\quad
\mu = \frac{1 + \sigma}{2 \sigma}\,.
\end{equation}
At small distances, potential $V(r)$ is either ill-defined or
complicated, but for $r \gg d$ it obeys the dipolar law $V(r) = 2
d^2 / r^3$ [Eq.~\eqref{eq:dipolar}]. From this, it is easy to see
that $\mu V(r) r^2 \ll 1$ at $r \gg b$ with $b$ given by
\begin{equation} \label{eq:b}
                         b = 8 \mu d^2\,.
\end{equation}
At such $r$ the potential energy $V$ acts as a small
perturbation.~\cite{Landau_III} Therefore, $\chi_m(r)$ coincides with the
wavefunction of a free particle,
\begin{equation} \label{eq:chi_far}
       \chi_m(r) = c_1 K_m(\varkappa r)\,,\quad r \gg b\,.
\end{equation}
Note that $b$ is either of the order or much larger than $d$ because $\mu \geq
2$ and $d \simeq d_c \sim 1$.

Sufficiently close to the critical $d$, we have $\varkappa \ll 1 / b$. In this
case there exists an interval of distances $b \ll r \ll b^{1/3} \varkappa^{-2 /
3}$ where we can drop the term $\varkappa^2$ in Eq.~\eqref{eq:chi_m} compared to
$\mu V(r)$. After this, Eq.~\eqref{eq:chi_m} admits the solution
\begin{equation} \label{eq:chi_m_mid}
\chi_m(r) = I_{2 m}\left(\sqrt{\frac{b}{r}} \,\, \right)
        - 4 c_2 K_{2 m}\left(\sqrt{\frac{b}{r}} \,\, \right)\,,
\end{equation}
where $I_{2 m}(z)$ and $K_{2 m}(z)$ are the modified Bessel function of
the first and the second kind, respectively.~\cite{Gradshteyn_Ryzhik}
The unit coefficient for $I_{2 m}(z)$ and the factor of $(-4)$ in front of
$c_2$ are chosen for the sake of convenience. The ground-state solution
is obtained for $m = 0$. Using the asymptotic
expansion~\cite{Gradshteyn_Ryzhik} of $I_{0}$ and $K_{0}$ in
Eqs.~\eqref{eq:chi_far} and \eqref{eq:chi_m_mid}, and demanding them to
be consistent with one another, we find for $m = 0$
and $b \ll r \ll \varkappa^{-1}$:
\begin{align}
\label{eq:chi_log}
\chi_0 &= 1 - 2 c_2 \left[\ln \left(\frac{4 r}{b}\right) - 2 \gamma
                         \right]
        + \mathcal{O}\left(\frac{b}{r}\right)\,,
\\
 \label{eq:c_2}
       c_2 &= -\frac{1}{6 \gamma + 2 \ln(b \varkappa / 8)}
           = \frac{1}{\ln (E_0 / E_B)}\,,
\end{align}
where
\begin{equation}
\label{eq:E_0}
E_{0} = \frac{8}{e^{6 \gamma}} \left(\frac{\sigma}{1 + \sigma}\right)^3
        \frac{1}{d^4}\,.
\end{equation}
Here $\gamma = 0.577\ldots$ is the Euler-Mascheroni
constant.~\cite{Gradshteyn_Ryzhik} Equation~\eqref{eq:chi_log} specifies the
boundary condition to which the solution for $\chi_0$ in the near field, $r
\lesssim b$, must be matched.

At $d = d_c$, both $\varkappa$ and $c_2$ vanish. Wavefunction
$\chi_0(r)$ at small $\varkappa$ can be viewed as the wavefunction
for $d = d_c$ perturbed by the small change in the boundary
condition in the far field, $r \gtrsim b$, and by another
perturbation,
\[
                \varkappa^2 + \left. \mu V(r) \right|^d_{d_c},
\]
in the near field, $r \lesssim b$. To the first order in these
perturbations we have
\begin{equation} \label{eq:E_B_pert_I}
               E_B = -A c_2 + B(d, \varkappa^2)\,,
\end{equation}
where $A$ is a constant and $B$ is a smooth function subject to the
condition $B(d_c, 0) = 0$.  Expanding $B$ to the first order in $d_c
- d$ and $\varkappa^2$, we arrive at the transcendental equation for
$E_B$:
\begin{equation} \label{eq:E_B_pert}
\left(1 - \mu \frac{\partial B}{\partial \varkappa^2}\right) E_B
+ \frac{A}{\ln (E_0 / E_B)} = -\frac{\partial B}{\partial d}\, (d_c - d)\,.
\end{equation}
The solution cannot be written in terms of elementary functions. However, the
logarithmic term gives the sharpest dependence on $E_B$. Hence, at small $E_B$ the
first term on the left-hand side of Eq.~\eqref{eq:E_B_pert} can be dropped. Now
this equation can be easily solved to recover Eq.~\eqref{eq:E_B_asym} with
\begin{equation} \label{eq:D}
                   D = \frac{A}{C}\,,\quad
                   C = -\frac{\partial B}{\partial d}\,.
\end{equation}
The coefficients $A$ and $C$ must be determined from the solution of
the inner problem. For $\sigma \ll 1$ part of this task can be
accomplished analytically, as explained later in this section. For
$\sigma \sim 1$ a numerical solution, such as the one discussed in
Sec.~\ref{sec:Numerical}, seems to be the only alternative.

Our results comply with a general theorem,~\cite{Bolle_84} which states
that in the asymptotic limit $k = i \varkappa \to 0$ the scattering
phase shift $\delta(k)$ satisfies the equation
\begin{equation} \label{eq:phase_shiftA}
(\pi/2) \cot \delta(k) = \ln(k / 2) + f(k^2)\,,
\end{equation}
where $f(z)$ is some analytic function. This theorem is valid
for a general short-range potential in 2D.
For a bound state $\cot \delta(k)$ should be replaced by $i$, leading to
\begin{equation} \label{eq:Bolle_result}
\ln \left( \sqrt{\mu E_B} / 2 \right) + f(-\mu E_B) = 0\,,
\end{equation}
which is in agreement with our Eq.~\eqref{eq:E_B_pert}. Our derivation
has the advantage of showing that the proper dimensionless combination
in the argument of the logarithm is $\sqrt{E_B / E_0}$ and that the
asymptotic behavior~\eqref{eq:E_B_asym} is realized at $E_B \ll E_0$.

\subsection{Binding energy for small mass ratios}

Although the electron-hole mass ratio is not truly small in typical
semiconductors, it is interesting to examine the case $\sigma \ll 1$
from the theoretical point of view. At such $\sigma$ the exciton
interaction potential $V$ can be meaningfully defined at all distances
using the Born-Oppenheimer approximation (BOA).~\cite{Born_Oppenheimer,
Born_51} In addition, the radial wavefunction can be computed everywhere
with accuracy $\mathcal{O}(\sigma)$.

The distance $r$ between excitons is no longer a physically
reasonable variable when the four particles approach each other
closely and their partitioning into excitons becomes ambiguous. In
the BOA this problem is mitigated by selecting $R$ --- the distance
between the heavy charges --- to be the radial coordinate of choice.
The ground-state biexciton wavefunction is taken to be
\begin{equation} \label{eq:Psi_BOA}
\Psi = \chi ({R}) \varphi (\mathbf{R}, \mathbf{r}_1, \mathbf{r}_2)\,,
\end{equation}
where $\varphi$ is the ground-state of two interacting electrons
subject to the potential of two holes fixed at positions
$\mathbf{R}_{1, 2} = \pm \mathbf{R} / 2$:
\begin{equation} \label{eq:H_BOA}
H_\text{BOA} \varphi = \left[-\nabla_1^2 - \nabla_2^2
                     + U_R(\mathbf{r}_1, \mathbf{r}_2)
                       \right] \varphi
                     = U_{\text{BOA}} \varphi\,.
\end{equation}
Here $U_{\text{BOA}}(R)$ is the corresponding energy. In turn, $\chi(R)$
is found from
\begin{equation} \label{eq:chi_BOA_I}
 -\frac{1 + \sigma}{R} \frac{d}{d R} R \frac{d \chi}{d R}
 + \mu \left[
U_{\text{BOA}}(R) - E_{\text{BOA}}
\right] \chi = 0\,.
\end{equation}
The BOA is known to have $\mathcal{O}(\sigma)$ accuracy. In principle, it can be
systematically improved.~\cite{Beyond_BOA} However, since below we will be
solving Eq.~\eqref{eq:chi_BOA_I} by means of the quasiclassical
approximation, which itself is known to be accurate only up to
$\mathcal{O}(\sigma)$, this is unwarranted.

Dropping all inessential $\mathcal{O}(\sigma)$ terms, we can simplify
Eq.~\eqref{eq:chi_BOA_I} as follows:
\begin{gather}
\label{eq:chi_BOA}
 -\frac{1}{R} \frac{d}{d R} R \frac{d \chi}{d R}
 + \left[ \varkappa^2 + \mu V(R) \right] \chi = 0\,,
\\
\label{eq:V_BOA}
V(R) \equiv U_{\text{BOA}}(R) - U_{\text{BOA}}(\infty)\,.
\end{gather}
Our task is to solve this equation with boundary conditions $|\chi(0)| < \infty$
at the origin and
\begin{equation} \label{eq:chi_mid}
\chi(R) \simeq I_{0}\left(\sqrt{\frac{b}{R}} \,\, \right)
        - 4 c_2 K_{0}\left(\sqrt{\frac{b}{R}} \,\, \right)
\end{equation}
at $b \ll R \ll b^{1/3} \varkappa^{-2 / 3}$, with $c_2$ given by
Eq.~\eqref{eq:c_2}.

\begin{figure}[b]
\centerline{
\includegraphics[width=3.0in]{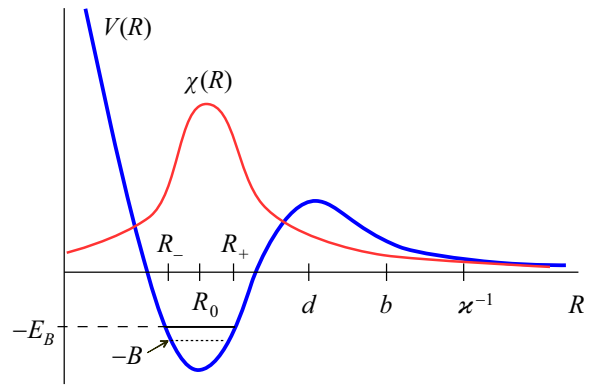}
}
\caption{Sketch of the interaction potential $V(R)$ and
the exciton wavefunction $\chi(R)$ for the Born-Oppenheimer
limit $\sigma \ll 1$.
\label{fig:WKB}}
\end{figure}

We reason as follows: in order to have a bound state, potential
$V(R)$ must be negative over some range of $R$. It can be shown that
this occurs in a single contiguous interval, see Fig.~\ref{fig:WKB}
and Sec.~\ref{sec:Numerical}. Inside of this interval there is a
classically allowed region, $\mu V(R) < -\varkappa^2$, where
function $\chi(R)$ reaches a maximum. As we approach the
dissociation threshold, this region shrinks. Near the threshold it
becomes very narrow, so that the quadratic approximation
\begin{equation} \label{eq:V_BOA_minimum}
\mu V(R) \simeq -\varkappa^2 + \frac12 \mu V^{\prime\prime}\,
                (R - R_-) (R - R_+)
\end{equation}
becomes legitimate. Here $R_-$ and $R_+$ are the turning points. To
construct the desired solution we simply need to match $\chi(R)$ in
the classical region, $R_- < R < R_+$, inside the tunneling region,
$R_+ \ll R \ll b$, and in the far field, $R \gg b$. Details of this
calculation are outlined in Appendix~\ref{sec:A}. The result is
\begin{align}
\label{eq:A}
A &= 4 \left(\frac{\pi}{e}\, \sigma V^{\prime\prime}\right)^{1/2}
       \exp (-2 S_0)\,,
\\
\label{eq:S_0}
S_0 &= \frac{1}{\sqrt{2 \sigma}}
    \int\limits_{R_+}^\infty {d R} \sqrt{V(R)}\,\,,
\\
\label{eq:B}
B &= |V(R_0)| - \sqrt{\sigma V^{\prime\prime}}\,,
\end{align}
where $R_0 = (R_+ + R_-) / 2$ is the point where $V(R)$ has the minimum.

%
%
%

Equations~\eqref{eq:D} and \eqref{eq:A} imply that the coefficient $D$ in
Eq.~\eqref{eq:E_B_asym}, and so the range~\eqref{eq:asym_range} of $d$ where
Eq.~\eqref{eq:E_B_asym} applies are proportional to the exponentially small
factor $e^{-2 S_0}$ at $\sigma \ll 1$. We expect that $D$ grows with $\sigma$
and by extrapolation, reaches a number of the order of unity at $\sigma \sim 1$.

A few other properties of function $d_c(\sigma)$ can also be deduced analyticaly.
For example, Eq.~\eqref{eq:B} implies that
\begin{equation}
 \label{eq:d_c_small_sigma}
       d_c(0) - d_c(\sigma) \propto \sqrt{\sigma}\,,
\quad \sigma \ll 1\,.
\end{equation}
Hence, $d_c(\sigma)$ has an infinite derivative at $\sigma = 0$ and so
initially decreases with $\sigma$. At some $\sigma$, however,
$d_c(\sigma)$ must start to increase. Indeed, due to the electron-hole
symmetry, the combination $d_c(\sigma) / (1 + \sigma) $ must have a
vanishing derivative~\cite{Adamowski_71} at $\sigma = 1$. Therefore,
\begin{equation}
 \label{eq:d_c_1}
       d_c^\prime(1) = d_c(1) / 2 > 0\,.
\end{equation}
Finally, we have a strict upper bound~\cite{Adamowski_71}
\begin{equation}
\label{eq:d_c_bound}
                 d_c(\sigma) \leq (1 + \sigma) d_c(0)\,.
\end{equation}
All of these properties are borne out by our Fig.~\ref{fig:d_c}. Still, a purely
analytical solution of the biexciton problem does not appear to be possible at
any $\sigma$. In the next section, we approach it by numerical calculations.

\section{Numerical simulations}
\label{sec:Numerical}

In order to verify our analytical predictions and other results in
the literature~\cite{Tan_05, Zimmermann_08}, we have carried out a
series of numerical calculations using the SVM. To implement this
method we customized the published SVM computer code~\cite{Varga_97}
for the problem at hand. In the SVM one adopts a nonorthogonal basis
of correlated Gaussians in the form~\cite{SVM_book}
\begin{equation} \label{eq:basissvm}
G_n = \exp
      \left(-\frac12 \mathbf{x}^\dagger \mathbf{B}_n \mathbf{x} \right)\,,
\end{equation}
from which a variational wavefunction of given electron and hole spins ($S$ and
$s$, respectively) is constructed:
\begin{equation} \label{eq:wavesvm}
\Psi = {\mathcal A} \left[G_n(\{ \mathbf{r}_\nu \})
                   \Upsilon_{S, s}\right]\,.
\end{equation}
Here $\mathbf{x}$ is a $3 \times 1$ vector of Jacobi coordinates
(linear combinations of differences in particle coordinates in which
the kinetic energy separates), $\mathbf{B}_n$ is a positive-definite
$3 \times 3$ matrix, $\mathcal{A}$ is the antisymmetrizer, and
$\Upsilon_{S, s}$ is the spin wavefunction. All our SVM calculations
are done for the spin-singlet state $S = s = 0$. Note that $G_n$
corresponds to the zero total momentum of the system.

%
%
%
\begin{figure}[t]
\centerline{
\includegraphics[width=3.0in]{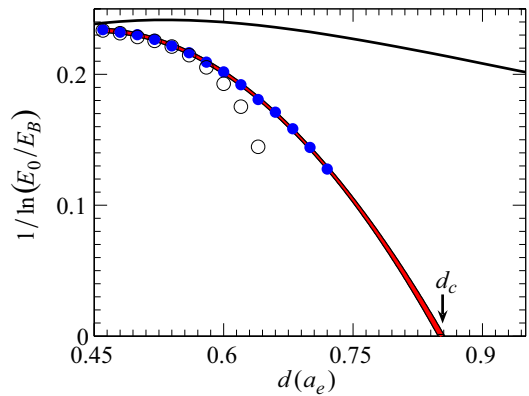}
}
\caption{Logarithmic plot of the biexciton binding energy as a
function of $d$ for $\sigma = 1$. Our results are shown by the
filled symbols; the open circles are from Ref.~\onlinecite{Tan_05}.
The thicker line is the fit to Eq.~\eqref{eq:log_E_B_asym}, which
yields $d_c = 0.87 \pm 0.01$ with a 95\% confidence level. The other
line is Eq.~\eqref{eq:Tan} with $\alpha$ and $\beta$ from
Ref.~\onlinecite{Tan_05}. \label{fig:fig5}}
\end{figure}

%
%
%
%
\begin{figure}[b]
\centerline{
\includegraphics[width=3.0in]{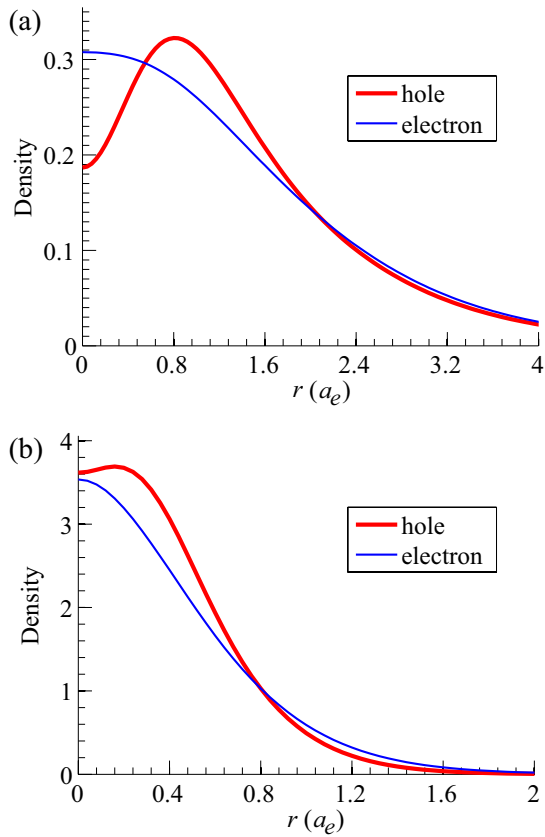}
}
\caption{(a) Electron and hole density \textit{vs\/}. the distance to the center
of mass in a biexciton with $\sigma = 0.5$ and $d = 0.0$.
(b) Same for $\sigma = 0.5$ and $d = 0.3$. \label{fig:density}
}
\end{figure}

%
%
%
\begin{figure}[t]
\centerline{
\includegraphics[width=3.0in]{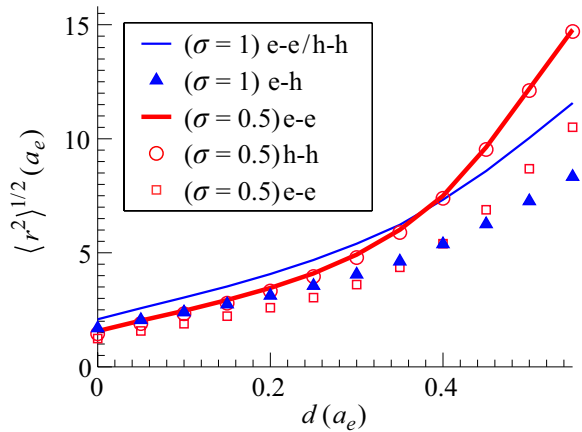}
}
\caption{Root mean square of the pairwise distances between the biexciton
constituents \textit{vs.\/} $d$ for $\sigma = 0.5$ and
$\sigma = 1$.
\label{fig:fig2}
}
\end{figure}

The number of basis states is grown incrementally until the energy
is converged or the prescribed basis dimension (typically $700$) is
reached. At each step a new quadratic form $\mathbf{B}_n$ is
generated randomly. If adding the corresponding function $G_n$ to
the basis improves the variational energy significantly, this $G_n$
is kept; otherwise, a new $\mathbf{B}_n$ is generated by varying
some of its matrix elements. Details can be found in
Refs.~\onlinecite{SVM_book} and \onlinecite{Varga_97}.

Our numerical results for $\sigma = 0.5$ and $\sigma = 1$ are given
in Table~\ref{table:E_B} and plotted in Fig.~\ref{fig:E_B}. In
Fig.~\ref{fig:fig5} we replot the binding energy $E_B$ for $\sigma =
1$ in a form suitable for testing Eq.~\eqref{eq:E_B_asym}:
\begin{equation}
 \label{eq:log_E_B_asym}
 \frac{1}{\ln (E_0 / E_B)} = \frac{d_c - d}{D} + \frac{(d_c - d)^2}{D_1}\,.
\end{equation}
Here we take into account one more term in the Taylor expansion of
the right-hand side of Eq.~\eqref{eq:E_B_pert_I} compared to
Eq.~\eqref{eq:E_B_pert}. Extrapolation of the data to $E_B = 0$
gives us $d_c$. The uncertainties in this parameter are estimated by
imposing a $95\%$ confidence level on the fit coefficients $d_c$,
$D$, and $D_1$. The same procedure has been applied to several other
mass ratios in the interval $0.1 < \sigma \leq 1$. The results for
$d_c$ are shown in Fig.~\ref{fig:d_c}. Their comparison with other
results in the literature will be addressed in
Sec.~\ref{sec:Discussion}.

At $\sigma \leq 0.1$ the range~\eqref{eq:asym_range} of $d$ where
Eq.~\eqref{eq:E_B_asym} applies is exponentially small. Even with
our highly accurate numerical method we were not able to probe this
range. Thus, we assumed that the nonanalytical correction $A
c_2(E_B)$ is undetectable on the background of $E_B$ in
Eq.~\eqref{eq:E_B_pert_I}, so that our numerical results for
$E_B(d)$ at such $\sigma$ are dominated by the regular contribution
\begin{equation} \label{eq:E_B_poly}
      E_B = C (d_c - d) + C_1 (d_c - d)^2 + \ldots
\end{equation}
Accordingly, at $\sigma \leq 0.1$ we deduced $d_c$ from the fit of
$E_B(d)$ to a quadratic polynomial. Additionally, we confirmed that
at $\sigma = 0.2$ the two fitting procedures give similar results:
$d_c = 0.59 \pm 0.01$ per Eq.~\eqref{eq:log_E_B_asym} \textit{vs\/}.
$d_c = 0.58 \pm 0.01$ per Eq.~\eqref{eq:E_B_poly}.

Finally, we have computed the electron and hole densities in the
biexciton as a function of their distance from the center of mass.
Examples are presented in Fig.~\ref{fig:density} for $d = 0.0$ and $d =
0.3$. In the latter case the particles are on average further away from
the center of mass. The same trend is also seen in the average root-mean
square separations between various particles, which are plotted in
Fig.~\ref{fig:fig2}. Their accelerated growth with $d$
occurs because the biexciton becomes less bound and eventually
dissociates.

\section{Discussion}
\label{sec:Discussion}

Let us compare our results with previous theoretical work. Early
studies of the biexcitons based on Hartree-Fock~\cite{Laikhtman_02}
or Heitler-London~\cite{Okumura_01} approximations provided initial
evidence for the existence of a finite threshold $d_c$ for the
biexciton dissociation. However, they gave a considerably lower
$d_c$ that we find here because these approximations did not account
for all correlation effects essential to the biexciton stability.

Comparing with more recent calculation~\cite{Tan_05} of the biexciton
binding energies by the DMC technique, we find an overall excellent
agreement. Still, our SVM occasionally slightly outperforms the DMC, see
Table~\ref{table:E_B}. Furthermore, in the SVM the estimate of the
ground-state energy decreases at each step, so that the statistical
noise is never an issue, unlike in the Monte-Carlo methods. Neither the
SVM nor the DMC is able to compute arbitrarily small binding energies;
therefore, in order to determine $d_c$, an extrapolation to $E_B = 0$ is
necessary. The clarification of what extrapolation formula should
be used for this purpose is an important finding of this work.
Equation~\eqref{eq:log_E_B_asym} represents the true asymptotic behavior
in the limit of small $E_B$ and indeed describes our numerical results
at such $E_B$ better than the interpolation formula~\eqref{eq:Tan}
plotted alongside for reference.

Another recent theoretical work on biexcitons used a
Born-Oppenheimer-like approximation to compute the threshold
interlayer separation $d_c(\sigma)$. At all $\sigma$ shown in
Fig.~\ref{fig:d_c}, our $d_c$ are higher than those reported in
Ref.~\onlinecite{Zimmermann_08}. The deviation is much larger than
the uncertainty of $d_c$ from our extrapolation procedure. This is
surprising because the adiabatic BOA~\cite{Comment_on_BOL} is known
to give a strict lower bound on the ground state
energy~\cite{Pack_70} while variational methods, such as our SVM,
give a strict upper bound. Since the energy of a single exciton
$E_X$ is usually computed extremely accurately, our binding energies
should be smaller than those of Ref.~\onlinecite{Zimmermann_08}.
Accordingly, their estimates of $d_c$ should exceed ours, not the
other way around. We suspect that the problem may again be related
to the manner in which the $E_B \to 0$ extrapolation was performed
in Ref.~\onlinecite{Zimmermann_08}. In any case, a significant
difference is seen only at $\sigma \sim 1$. At small mass ratios,
where the approximation of Ref.~\onlinecite{Zimmermann_08} becomes
accurate to the order $\mathcal{O}(\sigma)$, our results are in
better agreement.

Turning to the experimental implications of our theory,
observations of biexcitons in single quantum well systems have been
reported by many experimental groups.~\cite{Kleinman_82, Phillips_92,
Bar-Ad_92, Birkedal_96, Adachi_04, Kim_98, Langbein_02, Maute_03} In
contrast, no biexciton signatures have ever been detected in
electron-hole bilayers. Let us discuss how this can be understood based
on our results.

The first point to keep in mind is that the biexciton dissociation
threshold $d_c$ plotted in Fig.~\ref{fig:d_c} is a zero-temperature
quantity. For the biexcitons to be observable at finite
temperatures, $E_B$ must exceed $k T$ by some numerical factor. (As
usual in dissociation reactions,~\cite{Saha_21} this factor is
larger the smaller the exciton density is.) The coldest temperatures
demonstrated for the excitons in quantum wells are $T \sim
0.1\,\text{K}$ [Ref.~\onlinecite{Butov_01}]. The maximum separation
$d_*$ between the 2D electron and hole layers at which biexcitons
are still physically relevant in such structures can be roughly
estimated from
\begin{equation}
\label{eq:d_*}
                    E_B(d_*) = 10^{-3}\,\text{Ry}_e\,.
\end{equation}
Function $d_*(\sigma)$ is plotted by triangles in
Fig.~\ref{fig:d_c}. In GaAs quantum wells we have~\cite{Butov_00}
$\sigma \approx 0.5$, $a_e = 10\,\text{nm}$, and so $d_* \approx
4.5\,\text{nm}$. In comparison, the smallest center-to-center
separation that has been achieved in GaAs/AlGaAs and InGaAs/GaAs
quantum wells without compromising the sample quality is at least
twice as large.~\cite{Butov_07} Cold gases of indirect excitons have
also been demonstrated in AlAs/GaAs structures,~\cite{Butov_98} in
which $d$ is smaller, $d = 3.5\,\text{nm}$. But the electron Bohr
radius is also smaller, $a_e \approx 3\,\text{nm}$, so,
unfortunately, the dimensionless $d$ is about the same.

A more serious obstacle to the creation and observation of
biexcitons is disorder. A rough measure of disorder strength is
given by the linewidth of the exciton optical emissions, which is
currently $\sim 1\,\text{meV}$, i.e., of the order of
$0.1\,\text{Ry}_e$ in GaAs. $E_B$ becomes smaller than this energy
scale as soon as $d$ exceeds the thickness of a few atomic
monolayers, see Fig.~\ref{fig:E_B}. Actually, if the disorder were
due to a long-range random potential, it might still be possible to
circumvent its influence on the measured optical linewidth by
interferometric methods such as quantum beats.~\cite{Bar-Ad_92,
Adachi_04} In reality, a short-range random potential is probably
quite significant.

One potentially promising system for the study of the biexciton stability
diagram is a single wide quantum well subject to an external transverse
electric field.~\cite{Schultheis_87} If the well is symmetric and the
applied field is zero, we have $d = 0$. A finite field
can pull electrons and holes apart, leading to $d > 0$. Of course, for such
a structure one should recalculate the stability diagram of
Fig.~\ref{fig:d_c} by taking into account the motion of particles in all
three dimensions.

Although it is challenging to observe the binding of free indirect
excitons, in experiments they can be loaded and held together in
artifical traps.~\cite{Hammack_06} We anticipate that the SVM can be
a powerful tool to study systems of a few trapped excitons
theoretically, complementing recent Monte-Carlo
work.~\cite{Filinov_06}

In conclusion, we have obtained the most accurate estimates to date
of the binding energies of two-dimensional biexcitons. Future work
may include a refined study of exciton-exciton
scattering~\cite{Zimmermann_08} and interacting excitons in traps.


This work is supported by the NSF grant DMR-0706654. We are grateful to R.~Needs
for providing us with the numerical DMC results from Ref.~\onlinecite{Tan_05}
listed in Table~\ref{table:E_B}. We thank L.~Butov and L.~Sham for valuable
discussions.

\appendix

\section{Rigorous bounds for the biexciton binding energy}
\label{sec:Proof}

In this appendix we give a few strict upper bounds on $E_B$, which
enable us to prove the nonexistence of stable biexcitons at sufficiently
large $d$. The basic logic of the proof was outlined in
Sec.~\ref{sec:Analytical}A. Here we provide the technical details.

Our starting bound is
\begin{equation} \label{eq:E_B_vs_E_R}
 E_B \leq \max\limits_R E_R\,,
\end{equation}
where
\begin{equation} \label{eq:E_R}
E_R = \inf \text{spec}\, H_\infty - \inf \text{spec}\, H_R
\end{equation}
is the binding energy of the two-electron Hamiltonian $H_R = T_R
+ U_R$ whose kinetic term is
\begin{equation}
\label{eq:T_R}
T_R = -(1 + \sigma) (\nabla_1^2 + \nabla_2^2)\,,
\end{equation}
and the potential term $U_R$ is given by Eq.~\eqref{eq:U_R}. The Hamiltonian
$H_R$ is similar to that of the original problem
[Eqs.~\eqref{eq:T}--\eqref{eq:Phi}] except the holes are replaced by static
charges separated by a given distance $R$ and the electron mass is
made equal to the reduced electron-hole mass.

To derive the inequality~\eqref{eq:E_B_vs_E_R} we take advantage of the
well-known theorem that the ground-state energy as a concave function in
the strength of an arbitrary linear perturbation. (This theorem follows
from the variational principle.) For our purposes we choose the
perturbation in the form
\begin{equation}
\label{eq:T_new}
\Delta T_j = \nabla_j^2 - \left(\frac{d}{d \mathbf{R}_j}\right)^2\,.
\end{equation}
We add it to the kinetic energy terms with the coefficient $-\sigma \leq \tau
\leq 1$, yielding $T_j \to T_j + \tau \Delta T_j$.
Hamiltonians $H$ and $H_R$ are obtained by setting $\tau = 0$ and
$\tau = -\sigma$, respectively.

The perturbation leaves the reduced electron-hole mass invariant.
Therefore, it does not effect the ground-state energy $E_X$ of a
single exciton. The energy $E_\text{XX}(\tau)$ does vary with $\tau$
and the aforementioned concavity property dictates
\begin{equation}
\label{eq:E_XX_concavity}
E_\text{XX}(\tau) \geq \frac{1 - \tau}{1 + \sigma} E_\text{XX}(-\sigma) +
 \frac{\tau + \sigma}{1 + \sigma} E_\text{XX}(1)\,.
\end{equation}
Since $E_\text{XX}(-\sigma) = E_\text{XX}(1)$ by electron-hole symmetry, the
right-hand side is equal to $E_\text{XX}(-\sigma)$ for
all $\tau$. Consequently, $\tau = -\sigma$ gives the largest
binding energy and we arrive at the inequality~\eqref{eq:E_B_vs_E_R}.

If the kinetic energy $T_R$ is discarded, $E_R$ becomes equal to $-
V_{\text{cl}}(R, d) < 0$. We want to ascertain that quantum corrections
do not change the sign of $E_R$.

The quantum corrections appear in both $E_\text{X}$ and $E_\text{XX}$. The former are well
understood.~\cite{Lozovik_97} The internal dynamics of the exciton in the
large-$d$ case is analogous to that of a 2D harmonic oscillator with the
amplitude of the zero-point motion given by
\begin{equation}
\label{eq:r_eh_x}
\langle |\mathbf{r}_1 - \mathbf{R}_{1}|^2 \rangle = l^2\,,
\quad l = d^{3 / 4} (1 + \sigma)^{1 / 4} \ll d\,.
\end{equation}
The corresponding energy correction is
\begin{equation}
\label{eq:E_X_large_d}
E_X + \frac{2}{d} = \frac{2 \sqrt{1 + \sigma}}{d^{3 / 2}} - \mathcal{O}\left
(\frac{1}{d^2}\right)\,.
\end{equation}
This result immediately restricts the range of $R$ where the stable
biexciton may in principle exist. By positivity of the kinetic energy,
$E_R < 2 E_X - U_{\min}(R, d)$, where $U_{\min}$ is defined in
Sec.~\ref{sec:Analytical}A. Therefore, $E_R > 0$ may occur only at $R$
that satisfy
\begin{equation}
\label{eq:R_binding_min}
              V_{\text{cl}}(R) > 2 E_X + \frac{4}{d}\,.
\end{equation}
In view of Eqs.~\eqref{eq:V_cl} and \eqref{eq:E_X_large_d}, $R$ must
necessarily be much larger than $d$.

Choose an arbitrary $d_1$ such that $d \ll d_1 \ll R$. By definition of $U_{\min}$,
\begin{align}
\label{eq:interaction_bound}
U_R &\geq U_{\min}(R, d_1) + V_{Y}(\mathbf{r}_1) + V_{Y}(\mathbf{r}_2)\,,\\
\label{eq:V_Y}
V_{Y}(\mathbf{r}) &= \sum_{\mathbf{t} = \pm \mathbf{R} / 2}
[v(\mathbf{r} - \mathbf{t}, d_1) - v(\mathbf{r} - \mathbf{t}, d)]\,.
\end{align}
Accordingly, $E_R < 2 E_\text{X} - U_{\min}(R, d_1) - 2 E_{Y}$, where $E_{Y}$ is the
ground-state energy of a \textit{single\/} electron subject to the potential
$V_{Y}(\mathbf{r})$ of four out-of-plane charges. This potential has the shape
of two symmetric wells separated by the distance $R$. The amplitude of the
zero-point motion in each well is again $l \ll R$. Therefore, the energy shift due to
tunneling between the wells is exponentially small. (A rigorous upper bound
can be given.~\cite{Briet_89}) Furthermore, potential
$V_{Y}$ near the bottom of each well coincides with that of a single exciton up
to a constant
\begin{equation}
\label{eq:V_Y_shift}
\Delta V_{Y} = V_{Y}\left(\frac{\mathbf{R}}{2}\right)
             - \frac{2}{d}
             = \frac{2}{d_1} + \frac{d_1^2 - d^2}{R^3}\,.
\end{equation}
Hence, $E_{Y} = E_\text{X} + \Delta V_{Y}$ and
\begin{equation}
\label{eq:E_R_bound}
E_R \leq -\frac{2 d^2}{R^3}
 - \left[V_{\text{cl}}(R, d_1) - \frac{2 d_1^2}{R^3} \right]  \,.
\end{equation}
In these formulas we have dropped subleading terms
${o}({l^2}/{d_1^2})$, ${o}({d_1^4}/{R^5})$, \textit{etc\/}. With the
same accuracy the bracket in Eq.~\eqref{eq:E_R_bound} vanishes
[cf.~Eq.~\eqref{eq:V_cl}], so that we arrive at the result $E_R
\simeq -V_{\text{cl}}(R, d)$. This simply means that at large $d$
all quantum corrections to $E_R$ are parametrically smaller than the
direct dipolar repulsion of the two excitons. Therefore, $E_R \leq
0$ at all $R$, so that $E_{B} \leq 0$, and the proof is complete.

\section{Radial wavefunction for small mass ratios}
\label{sec:A}

In this appendix we show how the suitable solution of Eq.~\eqref{eq:chi_BOA} can
be constructed within the quasiclassical approximation. The necessary connection
formulas are derived by asymptotic matching with two exact solutions at small
and large $R$.

It is convenient to define the rescaled wavefunction $\psi(R) = \chi(R) \sqrt{R}$. From
Eq.~\eqref{eq:chi_BOA} we find that $\psi$ satisfies the equation
\begin{equation} \label{eq:psi}
\psi^{\prime\prime} -
\left(\varkappa^2 + \mu V(R) - \frac{1}{4 R^2}
\right) \psi = 0\,.
\end{equation}
This equation has two linearly independent quasiclassical solutions
\begin{equation} \label{eq:psi_WKB_I}
\psi_\pm(R) = \frac{1}{\sqrt{Q(R)}} \exp \big(\pm [S(R) - S(b)]\, \big)\,,
\end{equation}
where $Q$ and $S$ are given by
\begin{equation}
\label{eq:Q}
Q(R) = \sqrt{\varkappa^2 + \mu V(R)}\,,
\quad
S(R) = \int\limits_{R_+}^R d \rho Q(\rho)\,.
\end{equation}
The subtraction of the $R$-independent term $S(b)$ in the exponentials
amounts to multiplying $\psi_\pm$ by unimportant constants. This is
done purely for the sake of convenience. The reason for omitting the $1
/ 4 R^2$ term in the formula for $Q$ is more subtle. It is
explained in detail in Ref.~\onlinecite{Berry_73}.

In the following we assume that $\varkappa \ll 1 / b$, in which case there
exists a broad interval $d \ll R \ll b$ where potential $V(R)$ is dominated by
the dipolar repulsion~\eqref{eq:dipolar}. In this interval, $\mu V(R) \simeq b /
4 R^3 \gg \varkappa^2$; therefore,
\begin{equation} \label{eq:psi_WKB_II}
\psi_\pm(R) \simeq \left(\frac{4}{b} R^3\right)^{1/ 4}
 \exp \left[\pm \left(1 - \sqrt{\frac{b}{R}}\,\,\right)\right]\,.
\end{equation}
Using the asymptotic expansion formulas~\cite{Gradshteyn_Ryzhik} for
$I_0$ and $K_0$, it is easy to see that the following linear combination
\begin{equation} \label{eq:psi_WKB_III}
\psi(R) \simeq \frac{e}{2 \sqrt{\pi}} \psi_-(R)
        - \frac{2 \sqrt{\pi}}{e} c_2 \psi_+(R)
\end{equation}
of the quasiclassical wavefunctions~\eqref{eq:psi_WKB_II} smoothly matches with
the exact solution~\eqref{eq:chi_mid} at $d \ll R \ll b$. This is our first
connection formula. It is crucial for this derivation because in the
intermediate range of distances $b \ll R \ll \varkappa$ the quasiclassical
approximation breaks down. (It is invalidated by the sharp decrease of $V$ with
$R$.) In that region $\chi(R) = \psi / \sqrt{R}$ exhibits a slow logarithmic
falloff~\eqref{eq:chi_log} instead of the algebraic decay suggested by
Eq.~\eqref{eq:psi_WKB_II}. As explained in Sec.~\ref{sec:Analytical}, the
nonanalytical behavior~\eqref{eq:E_B_asym} of the binding energy
is precisely due to this logarithmic falloff.

To finish the calculation we need a second connection formula
between $\chi$ given by Eq.~\eqref{eq:psi_WKB_III} and the same
function near the classical turning point $R_+$. To find it we take
advantage of the exact solution for the harmonic oscillator
potential~\eqref{eq:V_BOA_minimum} in terms of the parabolic
cylinder function,~\cite{Gradshteyn_Ryzhik}
\begin{equation} \label{eq:psi_D}
\psi \propto D_{\varepsilon - 1/2}(-\sqrt{2}\, x)\,,
\quad x = \frac{R - R_0}{l}\,.
\end{equation}
Here $R_0 = (R_+ + R_-) / 2$ is the point where the potential $V(R)$ has
the minimum, $l = ({2} \,/\, {\mu V^{\prime\prime}} )^{1 / 4}$ is the
amplitude of zero-point motion about this minimum, and $\varepsilon$,
given by
\begin{equation} \label{eq:varepsilon}
\varepsilon = \frac12 l^2\, \big( \mu |V(R_0)| - \varkappa^2 \big)\,,
\end{equation}
is the corresponding energy in units of the oscillator frequency $\omega
= 2 \,/\, \mu l^2$. For the ground state we expect
\begin{equation} \label{eq:delta}
\delta \equiv \varepsilon - \frac12 \ll 1\,.
\end{equation}

The negative sign in the argument of $D_{\varepsilon - 1/2}$ in
Eq.~\eqref{eq:psi_D} is chosen to obtain an exponentially decaying wave
at large negative $x$, i.e., from the left turning point $R_-$ and
towards the origin. At large positive $x$, that is, at $R - R_+ \gg l$,
both decaying and growing exponentials are present. At such $x$ the
wavefunction can be cast into the quasiclassical form
\begin{equation} \label{eq:psi_WKB_IV}
\psi \simeq \sum_{\nu = \pm} \frac{c_\nu}{\sqrt{x}} \exp \left( \nu
\int\limits_{\sqrt{2 \varepsilon}}^x d \xi
                              \sqrt{\xi^2 - 2 \varepsilon}\,\,
\right)\,,
\end{equation}
which is equivalent to
\begin{equation} \label{eq:psi_WKB_V}
\sqrt{l}\,\, \psi(R) \simeq c_- e^{-S(b)} \psi_-(R)
                     + c_+ e^{S(b)} \psi_+(R)\,,
\end{equation}
see Eqs.~\eqref{eq:V_BOA_minimum}, \eqref{eq:psi_WKB_I}, and
\eqref{eq:psi_D}. This is our second connection formula
except we still have to specify the preexponential factors $c_+$ and $c_-$.
In fact, only their ratio is important. With the help of the
asymptotical expansion~\cite{Gradshteyn_Ryzhik} for $D_{\delta}$, one
finds it to be~\cite{Miller_53}
\begin{equation}
\label{eq:c_ratio_I}
\frac{c_+}{c_-} \simeq -2 \sqrt{\pi e}\, \delta\,.
\end{equation}
Comparing Eqs.~\eqref{eq:psi_WKB_III} and \eqref{eq:psi_WKB_V}, we obtain
\begin{equation} \label{eq:c_ratio_II}
\delta \simeq -\frac{1}{2 \sqrt{\pi e}}\, \frac{c_+}{c_-}
       \simeq 2 \sqrt{\frac{\pi}{e}}\, c_2 e^{-2 S(b) - 2}.
\end{equation}
For $\varkappa$ at which the above calculation is valid we have $S(b) \simeq S_0
- 1$, where $S_0 = S(R = \infty, \varkappa = 0)$. Thus, we arrive at
\begin{equation} \label{eq:varkappa_II}
\varkappa^2 \simeq \mu |V(R_0)| - \frac{1}{l^2}
- 4 \sqrt{\frac{\pi}{e}}\, \frac{c_2}{l^2} e^{-2 S_0}\,,
\end{equation}
which leads to Eqs.~\eqref{eq:A}--\eqref{eq:B} of
Sec.~\ref{sec:Analytical}.

Finally, a minor technical comment is in order. Since we have used
the quasiclassical approximation, all coefficients in
Eq.~\eqref{eq:varkappa_II} have a relative accuracy
$\mathcal{O}\left(e^{-S(b)} \right)$. In particular, we expect that
in place of $V(R_0)$ we have a slightly more negative value, so that
the ground-state energy $E_B$ never exceeds the oscillator
ground-state energy $V(R_0) + 1 / (2 \mu l^2)$, as required by
physical considerations.



\begin{thebibliography}{99}
\bibitem{Beyond_BOA} To this end one can iteratively diagonalize the
four-body biexciton Hamiltonian $H_\text{XX}$ by a sequence of canonical
transformations.~\cite{Bunker_77, Weigert_93} Equivalently, in the Lagrangian
formalism, one would integrate out two out of four fermion degrees of freedom.
This generates corrections to the potential and kinetic
terms of Eq.~\eqref{eq:chi_BOA}.
\bibitem{Comment_on_BOL} Actually, the approximation of
Ref.~\onlinecite{Zimmermann_08} differs from the usual
adiabatic BOA by using the exciton mass $M = m_e + m_h$ instead of the
heavy particle mass $m_e$, i.e., by replacing $1 + \sigma$ by unity
in Eq.~\eqref{eq:chi_BOA_I}. This decreases the kinetic energy
and thus lowers the ground state energy even further.
\end{thebibliography}
\end{document}